\documentclass[sigconf]{acmart}

\usepackage[utf8]{inputenc}

\begin{document}
\title{Challenges in IT Operations Management at a German University Chair -- Ten Years in Retrospect}

\renewcommand\footnotetextcopyrightpermission[1]{}
\settopmatter{printacmref=false}
\pagestyle{plain}

\author{Martin Geier}
\affiliation{\institution{{\small{}Chair of Real-Time Computer Systems}\\{\normalsize{}Technical University of Munich}}}
\email{geier@rcs.ei.tum.de}

\author{Samarjit Chakraborty}
\affiliation{\institution{{\small{}Chair of Real-Time Computer Systems}\\{\normalsize{}Technical University of Munich}}}
\email{samarjit@tum.de}

\renewcommand{\shortauthors}{M. Geier et al.}

\begin{abstract}
Over the last two decades, the majority of German universities adopted various characteristics of the prevailing North-American academic system, resulting in significant changes in several key areas that include, e.g., both teaching and research.\linebreak
The universities' internal organizational structures, however, still follow a traditional, decentralized scheme implementing an additional organizational level -- the Chair -- effectively a ``mini department'' with dedicated staff, budget and infrastructure.
Although the Technical University of Munich (TUM) has been establishing a more centralized scheme for many administrative tasks over the past decade, the transition from its distributed to a centralized information technology (IT) administration and infrastructure is still an ongoing process.
In case of the authors' chair, this migration so far included handing over all network-related operations to the joint compute center, consolidating the Chair's legacy server system in terms of both hardware architectures and operating systems and, lately, moving selected services to replacements operated by Department or University.
With requirements, individuals and organizations constantly shifting, this process, however, is neither close to completion nor particularly unique to TUM.
In this paper, we will thus share our experiences w.r.t. this IT migration as we believe both that many of the other German universities might be facing similar challenges and that, in the future, North-American universities -- currently not implementing the chair layer and instead relying on a centralized IT infrastructure -- could need a more decentralized solution.
\\
Hoping that both benefit from this journey, we thus present the design, commissioning and evolution of our infrastructure.
\end{abstract}

\maketitle

\section{Introduction}
\label{sec:intro}
With information technology (IT) pervading nearly all aspects of today's university life for both students and employees, IT operations management teams face an ever increasing number of challenges to ensure availability, security, applicability and\newpage\noindent usability of required and offered tools.
In case of \emph{research}, this includes services for knowledge dissemination and information sharing, computing and storage, provisioning of common software packages and -- as in all other cases -- user support.
\emph{Teaching} also relies on various IT-driven workflows for student lifecycle management and exam handling.
Depending on the university and the field of study, an increasing number of lab courses also heavily rely on specialized IT infrastructure.
Lastly, \emph{administration} such as human resources, accounting and facilities also depends on IT -- ranging from standard enterprise resource planning solutions to custom special tools.

Comparing the internal organizational structures of North-American universities with those found in Germany, one key difference stems from the additional organizational layer that German universities utilize -- the \emph{chair}, which is often also referred to as an \emph{institute}.
A collection of such chairs constitutes a department (such as that of Electrical or Mechanical Engineering).
The chairs could be viewed as ``mini departments'' with their own administrative and IT staff, budget, and IT infrastructure.
This results in a lot of flexibility, which is also necessary for the many practical laboratories offered in German universities, but comes with considerable overhead.

Over the past decade, the Technical University of Munich (TUM) has been establishing a more centralized scheme for many administrative tasks, ranging from non-technical (e.g., project management, financial and human resources) to the various technical areas of responsibilities.
Breaking with the traditional, decentralized scheme with the chairs maintaining their own administration and IT infrastructure, TUM's effort has been to reduce overhead, cut down on duplication and move the freed-up funding from non-academic or administrative positions to increasing the number of academic positions (such as Assistant Professors).
Towards this, TUM has been in particular heavily pushing towards more centralized services and IT infrastructures.
However, the necessary centralized alternatives require time to be set up and cannot provide the flexibility the chairs have traditionally been used to.
On one hand, such a transition is thus associated with a significant number of non-trivial challenges.
On the other hand, it might become difficult, if not impossible, to move labs and projects of many German universities to the envisioned centralized infrastructure common in North-American universities.
Hence, we believe that this situation requires considerable planning and introspection, and a realization of the trade-offs that are involved in \emph{centralized} versus \emph{decentralized} IT infrastructures in German universities, especially taking into account the kind of hands-on lectures and laboratories that are offered.

\noindent
\textbf{Goals of this paper:}
TUM's efforts are not unique in Germany and many other local universities are following on the same track.
In this paper, we outline our experiences with the (continuing) transition process from a chair-oriented to a more centralized administration, particularly focusing on IT services and infrastructure.
First, other German universities, who face similar challenges that we do, might benefit from our experiences and perceived challenges.
Second, we hope to get feedback from our North-American counterparts who have extensive experience with centralized IT operations at universities.
Third, the worldwide trend in teaching and learning has been steadily shifting from traditional classroom-oriented lecturing to self-learning using online courses and MOOCs (Massive Open Online Courses).
In order to adopt to this growing trend, it is becoming important to focus more on labs and hands-on projects that might help students to better ``digest'' their newfound and self-acquired knowledge.
Further, online courses cannot replace the value of hands-on experiments and projects that require physical, electrical and software infrastructure.
Hence, providing them will also help the universities to retain their value, in addition to meaningfully supplementing what students can learn online on their own.
Towards this, providing suitable IT support -- going far beyond web browsing, emails and backed-up storage -- that might be necessary for these labs and hands-on projects is of paramount importance.
We believe that, in the future, a suitable IT setup might lie somewhere in between the traditional chair-oriented decentralized system in Germany and a centralized North-American approach.
Hence, our experiences outlined in this paper might also benefit IT administrators and planners from American universities.
To characterize the various, often lab-related peculiarities and requirements that drive IT operations at a German university chair, we present the -- at the moment still mostly decentralized -- IT system deployed at the Chair of Real-Time Computer Systems~(RCS) and its design, introduction and ongoing evolution towards more centralized services over the last ten years in retrospect.

The remainder of this paper is organized as follows.
At first, Sec.~\ref{sec:entities} introduces all entities involved in IT operations at chair-level -- covering not only different types of staff members at RCS, but also external TUM units and the compute center.
Sec.~\ref{sec:start} reconstructs the initial state of the IT at the time both authors joined the RCS (approx. ten years ago) and motivates the derivation of requirements for a future IT infrastructure in Sec.~\ref{sec:reqs}.
Based thereon, Sec.~\ref{sec:design} presents original design together with selected implementation details of the system as initially introduced in 2012.
Sec.~\ref{sec:evo} not only summarizes our findings during both start-up and operation of our new infrastructure, but also covers the external developments and their impact on our local IT operations.
Sec.~\ref{sec:con} finally concludes this paper.

\section{The Entities}
\label{sec:entities}
Due to the highly federated structure both within and outside the University, our local IT operations not only involve several people at the RCS itself, but also extend towards both various other organizational units within TUM and the joint compute center of Munich's public universities (as our highest-level IT, high-performance computing and internet service provider).

\subsection{The Human Beings: Scientific, technical \& non-technical Staff}
\label{subsec:hbs}
Each chair -- implementing a ``mini department'' as introduced below in Sec.~\ref{subsec:ous} -- is headed by (at least) one \textbf{professor} with almost unrestricted control of scientific and administrative matters.
In case of RCS, the second author joined TUM as a professor in 2009 and had to head the Chair, without any prior experience in German universities.
His lack of proficiency with a decentralized administration and, in particular, his implicit assumption that IT infrastructure and services should be the concern of the University and need not have to be managed by individual professors on chair-level, posed some initial challenges for the IT operations management at the Chair.

Most day-to-day research and teaching activities, however, are handled by the scientific staff comprising up to dozens of full-time \textbf{research associates} pursuing their PhD degrees.
In contrast to other countries and -- primarily -- in engineering and computer science departments, they commonly enjoy full positions funded either from public sources (allocated to each chair) or by third parties such as, e.g., industry and (national or international) research foundations.
This sound financial position of a research associate (RA), however, comes at the price of various responsibilities that -- partially -- depend on the source of funding.
Generally, RAs are either committed to funded research projects or heavily involved in the chair's teaching activities (i.e., by giving tutorials for the professor's lectures or running entire labs) -- or both.
In addition, most RAs are responsible for some of the various administrative tasks covering HR, funding, IT operations and organization of teaching and project-related matters at chair-level.
During his time as an RA, the first author, as an example, has been involved in one industry- and several agency-funded research projects, designed two new lab courses whilst also in charge of one external lecture and, at times, another lab.
The single most time-consuming assignment, however, turned out to be taking over and maintaining IT operations of the Chair.
With various -- predominantly outdated -- systems in existence at the time both authors joined RCS, a smooth transition to an up-to-date infrastructure was imperative to not only reliably, but also securely continue research and teaching activities.

In case of RCS in 2018/19, one professor, ten RAs and three external guests are teaching a total of eight lectures (partially\linebreak including a tutorial), five laboratory courses and seminars.
They are supported by four \textbf{technical and non-technical staff members} in charge of purchasing, IT, electronics workshop, secretary's office and finances -- most of them, however, being assigned to part-time positions only.
Regularly serving as a gateway between the chair's researchers and the various organizational units within and outside TUM, they perform a vital interface function whilst maintaining a lot of flexibility regarding administrative and technical aspects at chair-level.

External to the chairs, numerous mostly non-scientific staff members comprise central services (e.g., library, IT, language and international centers) plus functional and administrative units such as HR, financial, controlling, facilities and legal.
In total, TUM currently has over 10,000 employees with approx. two thirds in scientific and a third in remaining positions~\cite{tum}.

\subsection{The Organizational Units: Compute Center, University, Faculty \& Chair}
\label{subsec:ous}
From an administrative perspective, IT operations at chair-level require coordination of and contributions from technical staff across multiple organization units as some services -- per administrative decision or technical necessity -- are exclusively handled by only one, single unit.
This, e.g., holds true for the various essential network services made available to Munich's universities by the Leibniz Supercomputing Centre (LRZ)~\cite{lrz}, which serves as a \textbf{joint compute center} and gateway to the German National Research and Education Network (DFN)~\cite{dfn}.
Effectively both acting as Internet Service Provider (ISP) that also maintains an IP backbone for over 180,000 devices and operating various IT services in addition to High-Performance Computing (HPC) systems, the LRZ provides the foundation for most of the IT in research facilities in and around Munich.

The LRZ' services relevant for university, faculty and chairs today extend far beyond networking (i.e., switch management, routing, upstream IP connectivity and basic services including DNS and DHCP).
Additionally, the compute center not only operates both global end-user services (such as Wi-Fi, VPN or video conferencing) and per-client -- i.e., TUM-only -- services (e.g., campus management system, directory services or wikis), but also offers backup, storage and file sharing in addition to virtualized firewalling and compute nodes on a project basis.

The \textbf{University} itself today also manages a vast number of services within its various internal organizational units.
As a part of TUM's corporate IT systems and services, for instance, the central information technology unit takes care of facilities such as various web-based portals and managed workstations.
Additionally, it maintains an (SAP-driven) enterprise resource planning solution and the central campus management system primarily covering student-, teaching- and resource-related matters -- with the latter based on CAMPUSonline, a solution developed at the TU Graz, which also has been introduced by various other universities in both Austria and Germany~\cite{Grzemski:2018:CCD:3235715.3235719}.

More specialized services are provided by dedicated teaching and library units and include not only e-learning platforms and document/website support, but also (internal and public) repositories and e-access systems for scientific data exchange.

The \textbf{Department} of Electrical and Computer Engineering (which RCS is part of) complements selected services offered by neither LRZ nor TUM -- with some now also used by other departments.
Apart from student-only IT facilities such as the faculty's roughly 100 Linux workstations (operated together with one of its chairs) and a course scheduler, the Department provides not only a web-based management tool for additional administrative (e.g., examination-related) workflows, but also services essential for the (often predominantly) Linux-driven infrastructure used at its nearly 30 chairs.
Today, this includes both NFS4 storage servers and a Puppet-based configuration management system crucial for a wide, consistent provisioning of Linux servers and clients deployed by faculty and its chairs.

In case of RCS, the \textbf{Chair} itself has a long legacy regarding local IT systems and operations.
Active in the area of process control computing since 1972 and renamed as ``Chair of Real-time Computer Systems'' in 1999, the RCS has not only used, but also researched numerous computer architectures running various operating systems in both IT and real-time contexts.
Although each chair, department and university has its own history w.r.t. IT infrastructure, the authors hope to use RCS and its IT as a meaningful reference case in following sections.

\section{The Starting Point}
\label{sec:start}
This section introduces both the IT infrastructure of 2010 and some challenges the first author faced to maintain operations.

\subsection{State of the RCS' IT}
At the time both authors joined the RCS, a significant number of \emph{network} components and their operations had already been successfully transferred to the LRZ.
This includes a structured cabling infrastructure with central switches (providing up to six Gigabit Ethernet ports to each office seating two RAs) and the Domain Name System~(DNS) servers for the --~externally visible~-- internet domains of the chair.
Apart from a dedicated project VLAN (Virtual LAN) already interfaced to a ``virtual firewall'' instance provided by the LRZ' Cisco FWSM blades, however, all other lower-level network services such as internal firewalling, NAT (Network Address Translation), DNS and DHCP (Dynamic Host Configuration Protocol) were mapped to own hosts.
In case of the firewall implementing an iptables-based packet filter between the external upstream (via LRZ) and the chair's internal network, no failovers were available.

On the \emph{server} side, a large variety of hardware architectures and operating systems were used.
Although the majority of services were mapped to Intel/AMD-based systems running Linux and Windows, various non-x86 servers (such as Alpha-, MIPS- and Sparc-based machines with their respective flavors of UNIX) were an integral part of the system, e.g., providing additional disk space via NFS (Network File System).
Again for historical reasons, the majority of servers, local switches and the -- one or other -- UPS (Uninterruptible Power Supply) did not follow the standard 19-inch, rack-mount form factor.
Instead, a multitude of desktop chassis were distributed across the server room's tables -- with a variety of cables underneath.

The \emph{client} systems were -- and still are -- a combination of Intel/AMD-based desktops and notebooks used for general-purpose computing and, due to the Chair's research on real-time, various (mostly PowerPC- and ARM-based) embedded systems running specialized operating systems such as eCos, FreeRTOS and Real-Time Linux.
In both cases, the individual RAs have full administrative access to maintain and adapt the particular system to their needs -- which regularly resulted in the setup of server software to compensate a lack of centrally offered solutions.
Some of these services were even permitted through the firewall, e.g., to make them accessible for students connecting from home or via the LRZ-operated Wi-Fi directly.
Besides that, a significant number of client systems still used static IP and DNS configurations instead of relying on DHCP.

In 2010, the RCS' internal network thus consisted of ten non-x86 and 15 Intel/AMD-based servers (with four running Windows), ten printers and approx. 150 clients.
Old databases report nearly 1100 (mostly inactive) users in over 200 groups.

\subsection{Technical Challenges}
Traditionally, several RAs plus one member of technical staff were handling IT operations at RCS.
When the authors joined the Chair, however, the number of RAs still contributing had already reduced to one.
With said RA leaving RCS less than six months later, the first author quickly became the primary person in charge for maintaining the operation of the existing system, handling the pending migration and supporting users.

The \emph{hardware}'s average age and variety resulted not only in an increasing number of wear-out failures, but also additional effort to understand -- and, at least temporarily, resolve -- both various quirks and a current outage on each server platform.
Similar to hard drive, memory and fan problems, a number of PSU (power supply unit) failures were also difficult to remedy due to missing spare parts on site or the general unavailability of suitable replacements.
The first author remembers several cases of planned and unplanned power cuts that resulted in more than one server requiring a new PSU and -- in rare cases only -- even new hard drives with a subsequent data restore.
A single district-wide blackout revealed that five power circuits were not balanced properly, causing blown fuses at power-on.
To make matter worse, only three servers were connected to UPSs initially -- leaving the remaining majority unprotected.

From a \emph{software} perspective, keeping the IT infrastructure running required a steeper learning curve -- not only regarding regular (i.e., unchanged) operation, but also for more common administration tasks including user or host management.
The variety of non-Intel/AMD hardware architectures implied a large number of Operating Systems (OSs) that needed special, dedicated knowledge -- such as Tru64, RISC/os and Solaris.
Such knowledge not only was needed for operations of a single host (e.g., adding a replaced hard drive to its array), but also to cope with the historical, often unspoken -- and sometimes bizarre -- dependencies between servers and services.
The first author is reminiscent of realizing that running Matlab on the Intel/AMD Linux hosts required one of the Solaris servers to be operational as it provided the required disk space via NFS.
Other challenges were an rsync job partially synchronizing the configuration of some Linux servers -- occasionally overwriting local changes made by those unaware -- and a stale, live copy of the Chair's primary DNS zone on a server in another state.
Day-to-day administrative workflows often required multiple manual changes in tools or files on more than one host.
This, e.g., held true for the management of users and groups, which required registration on both a Linux-based YP/NIS (Yellow Pages or Network Information Service) server and a Windows NT domain controller.
Similarly, new disk space was manually allocated, formatted and exported on the (NFS/CIFS) server and -- on two other hosts -- added to automounter tables and import scripts.
Internal DNS and DHCP services, however, were centrally provisioned from a single, custom database -- a fact that not only enabled redundancy using multiple servers, but also simplified migration (to an even more central source).

This combination of hardware- and software-related issues made maintaining operations challenging -- not helped by the fact that most hardware (including the firewall or file servers) and services (such as email) neither had failover solutions on standby nor were monitored methodically or comprehensively.
Hardware defects thus often were detected rather late -- and in need of immediate attention, which often required dedication far beyond normal working hours -- similar to the case of the (albeit rare) power cuts during or shortly before a weekend.

Temporarily transplanting crucial existing servers to newer hardware was considered during the migration -- but actually never implemented due to various incompatibilities between installed software and available replacement hardware such as, e.g., missing device drivers for storage or network controllers.

The \emph{security} of the old system was questionable, too -- not only due to the often outdated/unmaintained server software still in use, but also because several services were also exposed to the public internet.
In rare occasions, old server daemons even inhibited installing updated client software, as in case of a new release of Adobe's Acrobat Reader and a rusty version of the Samba server interfering due to a certain CIFS feature.
Furthermore, the centrally operated user workstations relied on a KNOPPIX-based live system that, due to its dependency on the testing and unstable repositories of Debian, could not be updated over longer periods -- resulting in outdated clients.

From a \emph{user} perspective, however, only one of above issues was directly visible and regularly addressed -- the often limited availability of the IT system or some of its services.
This also included two Windows clients used by the non-technical staff members of the Chair.
Even though one required a (relatively) time-consuming setup to access TUM's SAP, neither backups nor replacement systems were at hand -- causing an occasional flurry in case of failure.
A further hindrance was the historical setup for email services that combined a local IMAP (Internet Message Access Protocol) server to retrieve or store messages with the LRZ's SMTP (Simple Mail Transfer Protocol) service for transmission of outgoing email.
The former relied on mbox-based storage for each individual folder -- resulting in massive performance penalties when accessing mailboxes larger than the server's file system buffers, an effect particularly noticable when moving emails between folders.
The latter was reachable only from the LRZ's own networks or by using a VPN (Virtual Private Network) -- causing additional discomfort for multiple smartphone users as the required VPN client was not available on all platforms.
Additionally, neither SVN (Subversion) for revision control nor wikis to cooperate with external partners were provided -- regularly complicating research and teaching activities.
Several hardware-centric lab courses also relied on custom, non-central solutions for storage and computing that greatly varied in terms of reliability, proficiency and security.
Most labs suffered from (undocumented) tweaks of file system permissions, often far beyond the ``usual mishmash'' resulting from Unix' and Windows' incompatible semantics, whilst two even depended on their own, again outdated NFS server plus a custom kernel module to control the Motorola-based boards using a pre-JTAG (Joint Test Action Group) interface.
Lastly, one newer lab relied on a complex, distributed runtime driven by custom, camera-based tracking system running on its own server, again interfacing central servers and user workstations.

\section{The Requirements}
\label{sec:reqs}
With the IT system's \emph{availability}, \emph{security} and \emph{maintainability} severely degraded, the following requirements for an updated, hopefully sustainable infrastructure were identified mid-2010.

Own \emph{hardware} (if not avoided completely) should be set up both in a structured fashion -- using, e.g., 19-inch rack-mount power distribution units, UPSs, switches and servers -- and such that redundancy is achieved for each type of component, whilst keeping not only their total number but also the variety of models as low as possible.
Based on an up-to-date, common hardware platform with a current choice of operating systems, a more reliable and secure IT should also simplify operations, e.g., by increasing the availability of file servers and firewall.

To improve security on a \emph{network}-level, services should not only be kept up-to-date (e.g., using software with dependable migration policies), but also be split into publicly and (only) internally exposed groups to reduce the potential impact.
In addition to assigning those to separate networks, only secured protocols should be used.
The Chair's internal network should eventually only contain various clients and non-public servers.

Both \emph{software} environments such as operating systems and each individual service implementation should be as hardware-independent as possible to enable or at least simplify recovery, migration and upgrades.
Even if all components are (initially) purchased in pairs to achieve redundancy, a (future) combined lack of spare parts and increase of wear-out failures will result in a situation as in 2010 and greatly benefit from an improved hardware-software independence of such a new infrastructure.

To further improve both availability and maintainability, a combination of hardware and service monitoring with beyond-host configuration traceability will help mitigating the impact of hardware failures or human error.
A single, or even multiple, but yet central sources of dynamic (e.g., host-, authentication- and storage-relevant) and static configuration should not only simplify (automated) system monitoring and tracing, but also reduce the number of entry points necessary for day-to-day administrative workflows.
The introduction of standard tools and documented operating procedures to consistently manage the configuration should lower the barriers for additional RAs to contribute and take over -- with the central documentation repository supporting functional printouts for severe outages.

For \emph{users}, various services and features should be provided securely, reliably and efficiently.
This not only includes email (with support for smartphones and large, i.e., up to 10~GBytes mailboxes of some RCS members), but also globally reachable SVN and wiki services (with support for guest accounts), file and compute servers, centrally operated user workstations (as our unified and up-to-date solution for research and teaching), redundant clients for the non-technical staff members (mainly for office tools and SAP) and a comprehensive documentation.
The system also should use state-of-the-art security measures for sensitive (in particular personal and teaching-related) data and provide unified ACL (Access-Control List) templates that ensure sane file system permissions on project, lecture and lab volumes.
The latter should further benefit from configuration templates interfacing own servers to the central infrastructure, which do not require non-standard (e.g., root-only) methods for common RA activities such as account resets and template deployment.
Selected, centrally maintained software packages could reduce setup and storage overhead for labs and research.

\section{The Design}
\label{sec:design}

\begin{figure}
	\includegraphics[width=1.0\linewidth,clip=true,trim=7mm 8mm 7mm 8mm]{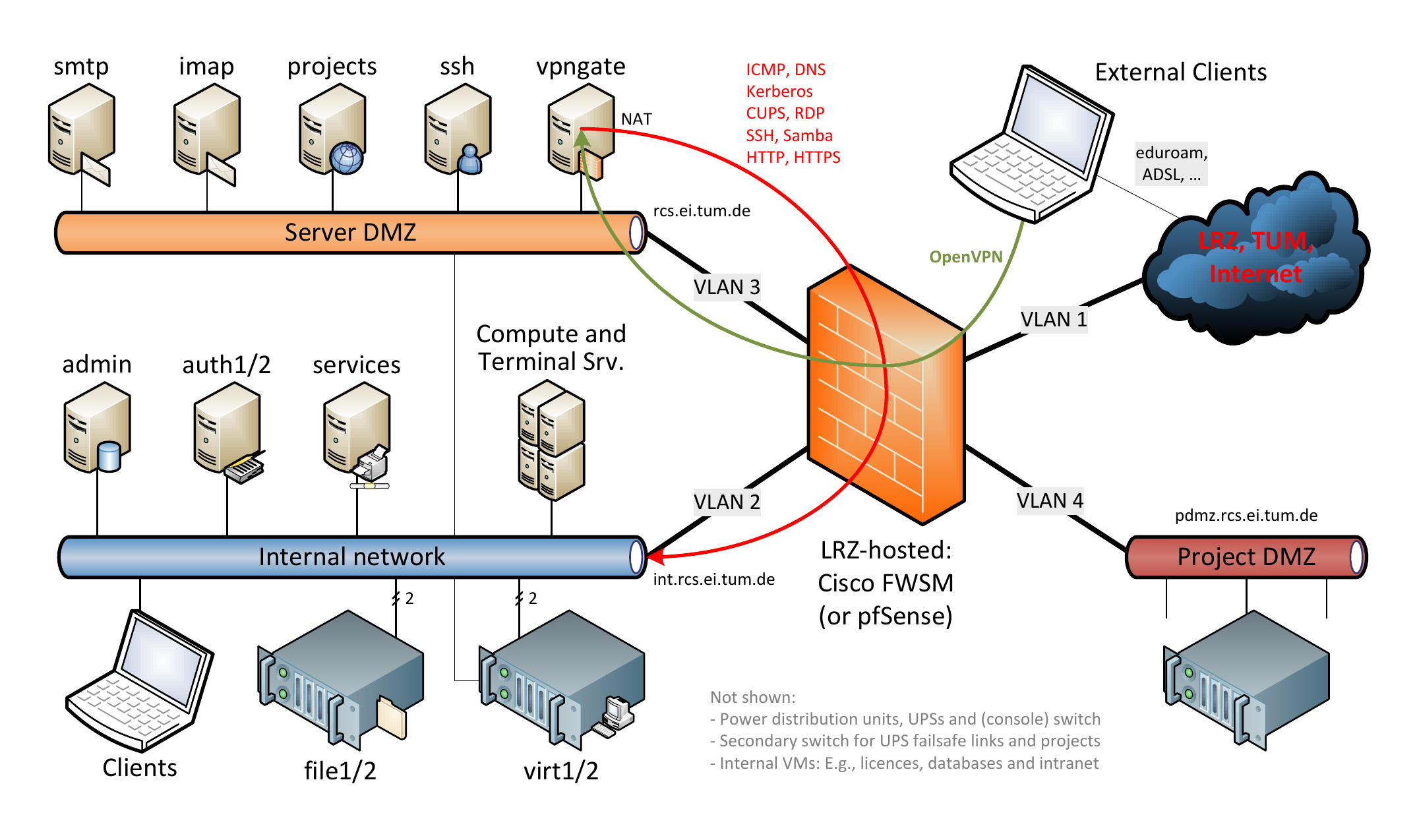}%
	\vspace{-1.3mm}
	\caption{Networks, Servers, Firewall and VPN-Gate}
	\label{fig1}
	\vspace{-1.3mm}
\end{figure}

The final hardware and software components were chosen and designed to incorporate redundancy, secure network protocols and configuration traceability -- throughout the entire system.

The \emph{hardware} was dramatically reduced to four new servers located in two 19-inch racks (conveniently donated by another chair) and complemented by three dedicated power circuits with corresponding distribution units and per-rack UPSs.
All servers feature redundant PSUs (on two independent circuits) and memory with support for Error-Correcting Code (ECC).
Each UPS is monitored by one server, which distributes status information to other hosts over redundant network paths to ensure a clean shutdown of all systems in case of power cuts.

The \emph{network} architecture was modified to not only provide one additional VLAN for globally reachable services, but also exclusively utilize a Cisco FWSM firewall offered by the LRZ.
Thus interfacing not only the internal RCS network, but also a server and a project DMZ (demilitarized zone) to the public internet, this firewall solution improves both availability (due to redundant hardware at LRZ) and security (as the internal network is no longer reachable from outside).
This topology is shown in Fig.~\ref{fig1}\footnote{The well-disposed reader might recognize the irony of documenting a mostly Linux-based IT infrastructure using Microsoft tools - The first author had not yet learned \textup{Ti\textit{k}Z} back then and thus resorted to Visio} and also reflected by separated DNS zones.

On the \emph{software} side, all four servers use Ubuntu Server as base OS.
Two identical quad-core Opterons with 8~GBytes of memory and 16-port RAID (Redundant Array of Independent Disks) controllers each are used as file servers (file1/2), whilst the other two feature two six-core Opteron CPUs, 32~GBytes of memory and 4-port RAID controllers each and serve as the virtualization hosts (virt1/2) for all services -- except storage.

We heavily utilize KVM (Kernel-based Virtual Machine), a hypervisor in current Linux kernels, to instantiate a dedicated hardware-independent VM (Virtual Machine) per ``group'' of services.
Individual VMs can be set up (using templates based on Ubuntu Server) and restored (from an rsync-based backup) within a few minutes.
This not only makes the complex service VMs independent of the underlying hardware (due to KVM's generic interface), but also enables a fast migration -- or even failover -- in case of failure on one of the virtualization hosts.

Three VMs implement a redundant DNS, DHCP and IAA (Identification, Authentication and Authorization) subsystem, which relies on a central LDAP (Lightweight Directory Access Protocol) directory for storage of host-, user- (including most passwords and email setup), group- and storage-related data.
Most information is managed using a web interface (originally developed by the City of Munich~\cite{limux}) only -- whilst hosts (also) and automounter tables (exclusively) are managed via custom command line tools.
Static host configuration data is centrally managed in Puppet, a configuration management tool using agents to ensure that all managed nodes and a central master are synchronized at all times.
With its class-based language, we implement a variety of host templates for, e.g., file servers, virtualization hosts, IAA VMs (admin and auth1/2, as shown in Fig.~\ref{fig2} bottom left), service VMs with and without IAA, our\linebreak ``basic Linux network client'' (with complete IAA and storage services) and a reduced version of the latter for (RA-operated) lab and project servers that still use central IAA and storage.

User IAA relies on multiple password hashes and Kerberos principals jointly stored in LDAP and integrated client-side using PAM (Pluggable Authentication Module) and GSSAPI (Generic Security Services Application Program Interface) on Linux, CIFS (Common Internet File System) with traditional Windows NT-like logons and, in both cases, Kerberos.
Storage is provided using NFS4 (Network File System version 4) using Kerberos and password-based CIFS.
This entire subsystem is centrally managed from a -- single -- configuration file, which not only configures file servers and clients as needed, but also controls on- and off-site backups of both user data and VMs.

Servers, VMs and services are monitored via Munin, whilst configuration is traced both locally (etckeeper and listchanges) and globally (using SVN repositories for LDAP and Puppet).

Most network protocols are secured either internally and by design (as Kerberos) or configured to enforce TLS (Transport Layer Security) -- with the only notable exception being CIFS.
A VPN gateway enables staff members to use selected services even when not connected to the internal network -- e.g., when using the LRZ's Wi-Fi on notebooks or working outside RCS.

Additional user \emph{services} include comprehensive email with fast, maildir-based IMAP, authenticated SMTP (with second password) and sieve filtering.
A projects VM provides global SVN and wiki services with fixed, random passwords whilst an Ubuntu-based diskless image drives central user workstations, also used to access the Terminal Server of non-technical staff.

The entire architecture of the Chair's new IT infrastructure is shown in Fig.~\ref{fig2} with VM and service details in~the~\mbox{appendix.}

\section{The Evolution}
\label{sec:evo}
To ensure a smooth \emph{start-up}, the system was tested during one semester using a newly created lab that required most services and the central user workstations -- with some of the centrally maintained software packages and USB firmware for its JTAG interface.
Step-by-step, labs were migrated successfully -- with only the most complex one with its distributed runtime and camera tracker posing a challenge.
The lab's server was linked to the central IAA and file servers, which ensured a consistent login and execution environment for all software components, plus SSO (Single Sign-On) for users.
After some initial hiccups due to ACL limitations and some fragile shell scripts, the lab eventually went live.
Shortly after the migration, the system already served over 370~users (150~lab accounts) on 200~clients,\linebreak handling 42~GBytes in mailboxes and over 20k mails a month.

Apart from three major outages of the primary fileservers due to an eventually fixed bug in a RAID controller's firmware, \emph{operation} has been smoothly.
With the introduction of newer Ubuntu releases, it became clear that our initial Puppet code-base requires restructuring, too~\cite{Plummer:2016:PII:2974927.2974950}.
Whilst missing support for SSH public key authentication was less critical than expected thanks to VPN-based Kerberos and fixed (project) passwords, all ACLs remain challenging due to incompatible applications.

After the migration, various \emph{external developments} affected design and use of the system.
With email, room booking and wikis now offered by TUM, both intranet and email VMs are no longer required.
The Department's file servers now offer NFS based on TUM's central IAA services, whilst the LRZ's gitlab service will eventually replace our SVN/wiki solution.

\section{Conclusion}
\label{sec:con}
In this paper, we introduced a key difference in organizational structures of German universities, which has resulted in rather decentralized IT operations at many chairs.
We presented the history, analysis and redesign of our Chair's infrastructure to share our findings -- in particular related to the various labs, which require specialized IT infrastructure.
With future, less (de?)central solutions in sight, IT operations remain exciting.

\appendix

\begin{figure*}
	\includegraphics[angle=0,width=1.0\linewidth,clip=true,trim=6mm 6mm 6mm 6mm]{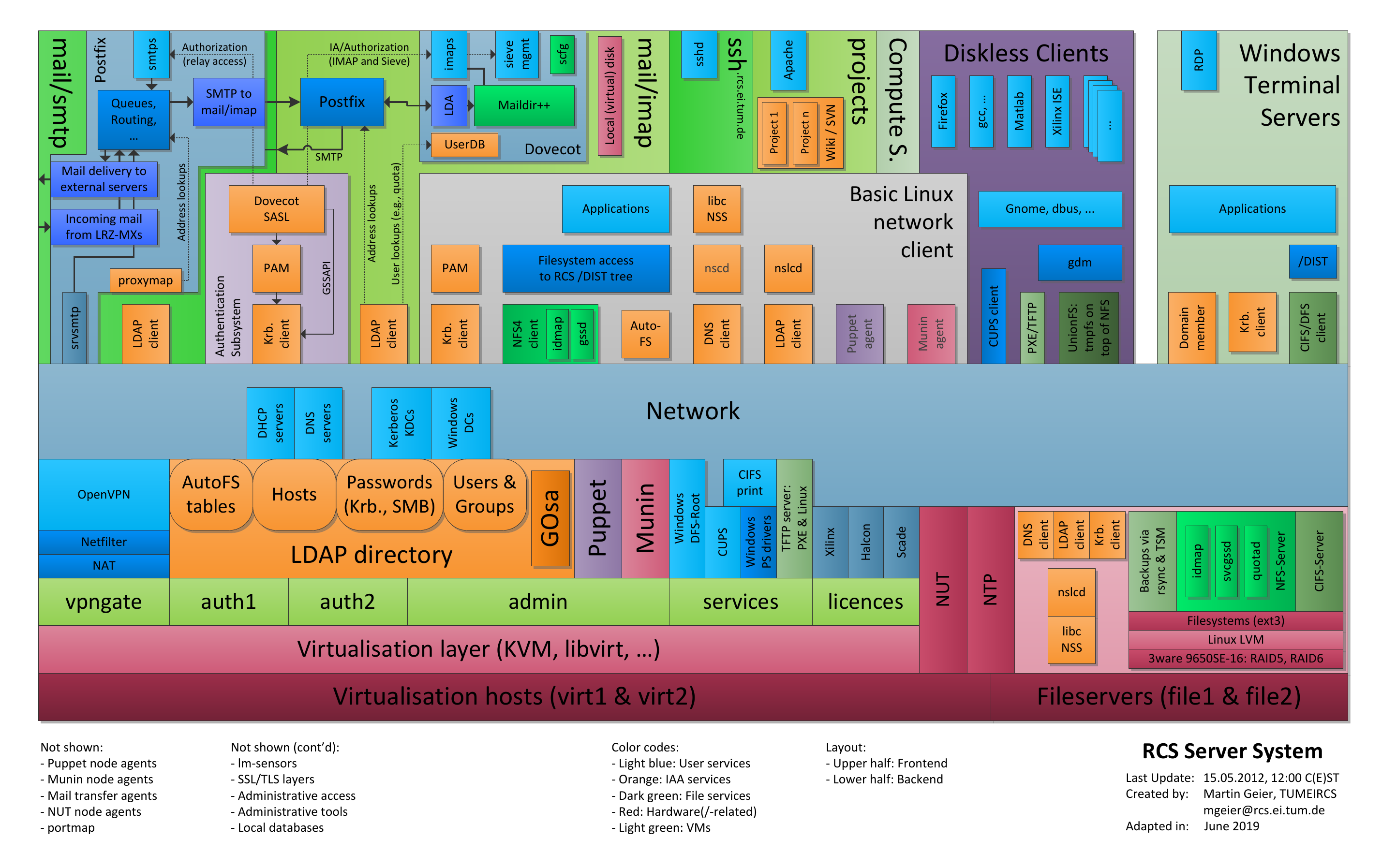}%
	\vspace{-1.7mm}
	\caption{RCS Infrastructure -- Physical Servers (bottom), Service VMs (light green) and User Frontend (top)}
	\label{fig2}
	\vspace{-1.7mm}
\end{figure*}

\section*{Backend Hosts and Service VM\MakeLowercase{s}}
Whilst all \emph{physical servers} use NTP (Network Time Protocol) daemons for time synchronization, NUT (Network UPS Tools) servers are only required on virt1/2 and forward UPS status to NUT clients on file1/2 and user VMs.
Puppet and Munin node agents are installed on all physical servers and VMs -- with the\linebreak same holding true for a minimal MTA (Mail Transfer Agent).
A Munin master (admin) captures physical (e.g., temperature, disk or RAID status) and logical (e.g., load or volume usage) samples and sends an email notification if limits are violated.

Both \emph{file servers} -- like many VMs -- use NSS (Name Service Switch) to access user and group information in LDAP, whilst GSSAPI enables NFS4 authentication using Kerberos.
Linux clients may choose between CIFS (implemented by the Samba server) and NFS4 for storage -- Windows ones only the former.
Disk space is organized using volumes, i.e., as individual ext3 file systems above RAID and LVM (Logical Volume Manager) exported via CIFS and kernel-based NFS.
Client-side imports rely on a Windows DFS (Distributed File System) entry point on the services VM and automounter tables in LDAP, whilst on- and off-site backups are implemented with rsync (between file1/2) and LRZ's Tivoli.
A variety of POSIX ACL templates and online mapping from CIFS/NFS4 to POSIX ACLs ensure a -- relatively -- consistent view and control of file permissions.

The \emph{LDAP directory} is managed with scripts and GOsa²~\cite{limux}, also storing Kerberos user principals with multiple passwords.

\section*{Frontend Services and Clients}
Most frontend (i.e., user-visible) services are provided by VMs,\linebreak e.g., services (for DFS, PXE, CUPS and Windows PS drivers).

Our ``basic Linux network client'' (center of Fig.~\ref{fig2}) serves as the foundation for several systems.
Its DNS and LDAP clients provide host, user and group name resolution over NSS, whilst Kerberos completes IAA with password-based login and (once authenticated) SSO to file, mail, SSH and web servers of RCS,\linebreak which also enables unified access to storage (via automounter).

SVN and wiki services are provided by the projects VM -- in the Server DMZ for global access.
Our Kerberos-authenticated web interface enables RAs to create projects and add(/remove) users, which receive random passwords (as the latter are often stored in plaintext).
Back-ends are located on an NFS~volume.

External and internal compute VMs provide SSH access for\linebreak users connecting from outside or executing complex jobs.
This also enables file synchronization and backups, e.g., via unison.

The centrally operated user workstations also rely on abovementioned client and provide an entire Ubuntu-based desktop environment booted via the network.
A variety of open-source and commercial applications, a flexible shell environment and SSO to central/custom servers simplify research and teaching.

Single-user Windows VMs and Terminal Servers offer many applications such as Office or certain, usage-restricted tools.
A dedicated set of VMs provides all tools (such as SAP) required by non-technical staff (connecting from central workstations).

\end{document}